\documentclass[prb,twocolumn,preprintnumbers,amsmath,amssymb,superscriptaddress,showpacs]{revtex4}
\usepackage{graphicx}
\usepackage{latexsym}
\usepackage{amsmath}
\usepackage{amssymb}
\usepackage{layout}
\usepackage{verbatim}
\usepackage{amsfonts,epsfig}
\usepackage[T1]{fontenc} 
\usepackage[cp1250]{inputenc} 

\newcommand{\beq}{\begin{equation}}
\newcommand{\eeq}{\end{equation}}

\newcommand{\CdMnTe}{Cd$_{1-x}$Mn$_{x}$Te}
\newcommand{\CdZnTe}{Cd$_{1-y}$Zn$_{y}$Te}
\newcommand{\ZnMnTe}{Zn$_{1-x}$Mn$_{x}$Te}

\begin{document}

\title{Magnetic polaron formation and exciton spin relaxation \\ in single CdMnTe quantum dots}
\author{{\L}. K{\l}opotowski} \email{lukasz.klopotowski@ifpan.edu.pl}
\affiliation{Institute of Physics, Polish Academy of Sciences, al.~Lotnik{\'o}w 32/46, 02-668 Warsaw, Poland}
\author{{\L}. Cywi{\'n}ski} 
\affiliation{Institute of Physics, Polish Academy of Sciences, al.~Lotnik{\'o}w 32/46, 02-668 Warsaw, Poland}
\author{P. Wojnar}
\affiliation{Institute of Physics, Polish Academy of Sciences, al.~Lotnik{\'o}w 32/46, 02-668 Warsaw, Poland}
\author{V. Voliotis}
\affiliation{Institut des NanoSciences de Paris, Universit{\'e} Pierre et Marie Curie, CNRS, 4 place Jussieu, 75252 Paris, France}
\author{K. Fronc}
\affiliation{Institute of Physics, Polish Academy of Sciences, al.~Lotnik{\'o}w 32/46, 02-668 Warsaw, Poland}
\author{T. Kazimierczuk}
\affiliation{Faculty of Physics, University of Warsaw, ul. Hoża 69, 00-681 Warsaw, Poland}
\author{A. Golnik}
\affiliation{Faculty of Physics, University of Warsaw, ul. Hoża 69, 00-681 Warsaw, Poland}
\author{M. Ravarro}
\affiliation{Institut des NanoSciences de Paris, Universit{\'e} Pierre et Marie Curie, CNRS, 4 place Jussieu, 75252 Paris, France}
\author{R. Grousson}
\affiliation{Institut des NanoSciences de Paris, Universit{\'e} Pierre et Marie Curie, CNRS, 4 place Jussieu, 75252 Paris, France}
\author{G. Karczewski}
\affiliation{Institute of Physics, Polish Academy of Sciences, al.~Lotnik{\'o}w 32/46, 02-668 Warsaw, Poland}
\author{T. Wojtowicz}
\affiliation{Institute of Physics, Polish Academy of Sciences, al.~Lotnik{\'o}w 32/46, 02-668 Warsaw, Poland}
\date{\today }

\begin{abstract}
We study the formation dynamics of a spontaneous ferromagnetic order in single self-assembled \CdMnTe\ quantum dots. By measuring time-resolved photoluminescence, we determine the formation times for QDs with Mn ion contents $x$ varying from 0.01 to 0.2. At low $x$ these times are orders of magnitude longer than exciton spin relaxation times evaluated from the decay of photoluminescence circular polarization. This allows us to conclude that the direction of the spontaneous magnetization is determined by a momentary Mn spin fluctuation rather than resulting from an optical orientation. At higher $x$, the formation times are of the same order of magnitude as found in previous studies on higher dimensional systems. We also find that the exciton spin relaxation accelerates with increasing Mn concentration.

\end{abstract}


\maketitle


Doping semiconductor quantum dots (QDs) with magnetic ions offers a possibility of controlling magnetic properties of matter at nanoscale. Notably, several theoretical reports have proposed tailoring of QD magnetization by tuning the number of carriers in a dot.\cite{fer04,gov05,abo07} However, in order to achieve the control over magnetization a detailed knowledge of its dynamics is required.
In compound II -- VI QDs the Mn doping is performed routinely enabling studies of very dilute systems including QDs with single Mn ions \cite{bes04} and of highly doped ones with molar contents up to 7\%.\cite{mak00} Magnetic properties are comfortably monitored through optical experiments, since exchange interaction between the localized magnetic ions and the band carriers leads to pronounced magnetooptical effects.\cite{dob03} In particular, energy minimization of a complex consisting of a photocreated electron-hole pair (an exciton) interacting with Mn ions, results in a spontaneous formation of a local ferromagnetic order -- a magnetic polaron (MP).

Static and dynamic properties of MPs have been subject to intensive experimental and theoretical studies\cite{naw81,die82,gol83,har83,zay87,mac94,mer95,mak00,seu02,woj07,mac04,gur08,bea09,sel10} Experimental fingerprint of the MP formation is a redshift of the exciton photoluminescence (PL) by polaron energy $E_{P}$ -- the energy gained by formation of the ferromagnetic order. The development of the magnetization can therefore be monitored in a time-resolved (TR) PL experiment, in which a transient shift of the exciton energy is observed allowing to evaluate the MP formation time, $\tau_f$. \cite{har83,zay87} However, in bulk and 2D systems a prerequisite for the MP formation is an initial localization of the exciton.\cite{mac94} A precise experimental identification of $E_{P}$ and $\tau_f$ is then hindered by processes related to trapping of the exciton. On the other hand, excitons in QDs are inherently localized by the QD potential, and thus the studies of MP formation dynamics in these nanostructures are free of the obscuring localization effects.\cite{seu02,bea09} The studies reported so far were performed on QD ensembles, in which the obtained $\tau_f$ may be inaccurate due to inhomogeneities in dot morphology leading to variations in exciton lifetimes,\cite{mac04red} $\tau_{X}$, affecting in turn the TRPL transients.
Previous reports have also left an open question regarding the mechanism of the MP formation. Specifically, it is important to assess whether an optical orientation of MPs is possible as suggested by quasiresonant PL studies \cite{mac04} or whether the MP orientation is determined by the direction of the magnetization fluctuation at the instant of exciton capture,\cite{gur08} as it was shown\cite{mer95} to occur in higher-dimensional systems.

Here, we investigate polaron formation in {\em single} \CdMnTe\ QDs. We study the formation dynamics for dots containing from $x \! = \! 0.01$ to $0.2$ molar fraction of Mn cations, extending the range of reported concentrations. For dilute samples, we find that the MP formation is interrupted by exciton recombination and we estimate the MP formation time $\tau_{f}$ on the order of a few nanoseconds. A different situation is encountered for $x \! \geq \! 0.1$, where we observe a distinct temporal redshift of the exciton energy with the equilibrium reached on a timescale of $\sim$100 ps. In order to assess the MP formation mechanism, we study the exciton spin relaxation dynamics by measuring the PL circular polarization. We find that the exciton spin is relaxed before the MP is formed and therefore we conclude that optical orientation of polarons is questionable.

Two sets of samples were grown. The first consisted of QDs with $x\!<\! 0.035$, the second featured dots with $x\! > \! 0.05$. Both sets were grown by molecular beam epitaxy on a semi-insulating, (001)-oriented GaAs substrate with a $\sim \!4$ $\mu$m thick CdTe buffer layer. In the first set, the buffer was followed by a \CdZnTe\ layer ($y\! \approx \! 0.7$) providing a necessary lattice mismatch for the dot formation and shifting the QD PL band below the intrashell Mn transition.\cite{seu02} Next, a single layer of dots was formed from a strained layer of 2 nm thick \CdMnTe. The dots were then covered with a 50 nm thick \CdZnTe\ barrier layer. Using this approach to produce QDs with $x \!> \! 0.05$ resulted in dots with probable type-II band alignment, evidenced by fourfold increase in exciton lifetime $\tau_{X}$ and lack of any signatures of the MP formation. In order to maintain the type-I band alignment in QDs with $x\! > \! 0.05$, we followed the growth of a CdTe buffer with depositing a layer of \ZnMnTe\ with $x$ matched to the Mn content in the dots. The resulting samples exhibited analogous $\tau_{X}$ to the low $x$ ones and temperature dependence of PL showed signatures of the MP formation. In order to access single dots, 500 nm shadow mask apertures were processed by spin casting polybeads before deposition of a 200 nm thick gold layer.

Polaron formation dynamics was studied by TRPL at $T \! = \! 8$ K. In the case of low $x$ samples, the PL signal was excited with a frequency doubled output of an optical parametric oscillator pumped with a 2 ps Ti:sapphire laser. Excitation energy was tuned below barrier band gap, to 2.19 eV, into resonance with a high energy tail of the inhomogenously broadened PL band. These excitation conditions assured negligible heating of the Mn spin system since no carriers in extended excited states were created.\cite{cle10} In the case of samples with high $x$, this excitation mechanism was inefficient due to a pronounced intrashell Mn transition.\cite{seu02} Therefore, in order to increase absorption, we excited the PL within the barrier band gap, with a frequency doubled output of the Ti:sapphire laser, at 3.1 eV. In this case, the heating was suppressed by an efficient spin-lattice relaxation characteristic for high $x$ alloys.\cite{die95} PL signal was detected and time-resolved with a streak camera. To retrieve transient PL spectra, we integrated 10 ps intervals of the streak images. The overall temporal resolution of the detection setup was better than 15 ps.


\begin{figure}[!h]
  \includegraphics[angle=0,width=.5\textwidth]{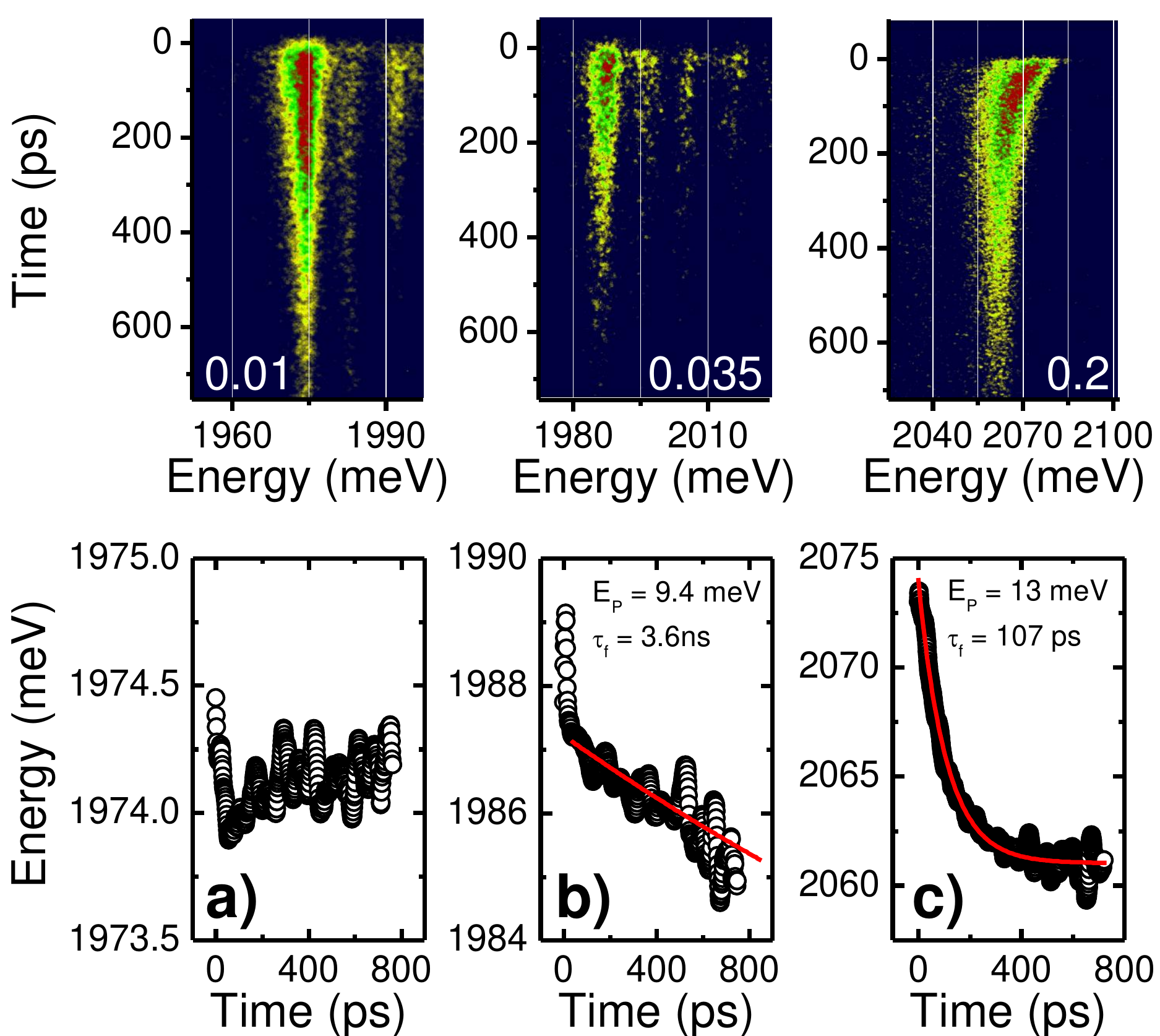}
 \caption{First row: Streak camera images of three QDs with $x=$ 1\%, 3.5\%, and 20\%. Second row: Corresponding exciton energies as a function of time obtained by fitting the transition line with a Gaussian. }
  \label{raw}
\end{figure}

In order to calculate the static properties of the MP it suffices\cite{die82,gol83} to replace the individual Mn spins by a collective spin $\mathbf{\hat{S}}$, and write the Mn-exciton interaction as $\hat{H} \! = \! -(\alpha\mathbf{\hat{S}}\cdot\mathbf{\hat{s}}+\beta\hat{S}_{z}\hat{j}_{z}/3)/V$, where $\alpha$ and $\beta$ are the conduction and valence band exchange constants (in \CdMnTe\ $N_0\alpha \! = \! 0.22$ eV and $N_0\beta \! = \! -0.88$ eV, with $N_{0}$ being the cation density), $\mathbf{\hat{s}}$ and $\mathbf{\hat{j}}$ are the electron and hole spins, and the localization volume $V \! =\! [\int |\Psi|^4(\mbox{\bf r}) d^3 (\mbox{\bf r})]^{-1}$, where the exciton wavefunction $\Psi$ is crudely assumed as the same as for the electron and the hole. The exciton acts upon the Mn spins with an exchange field, which for a rectangular wavefunction is given by\cite{kav99} $B_{\text{X}} \! \approx \!  (\beta-\alpha) / (4\mu_B V)$, where $\mu_B$ is the Bohr magneton. In the linear response regime, when $B_{\text{X}}$ is smaller than the saturation field, the MP energy reads\cite{die82} $E_{P} \! = \! V\chi B_{\text{X}}^2$, where $\chi$ is the Mn susceptibility.

In Figure 1a, we present a streak image (top) and corresponding exciton energy positions (bottom) for a QD with $x\! = \! 0.01$. The linewidth of this transition, $\gamma \! = \! 3.6$ meV, is more than an order of magnitude larger than for nonmagnetic CdTe QDs.\cite{bac02,kaz09} This broadening occurs due to exchange interaction between the exciton and fluctuating Mn spins.\cite{bac02} The resulting linewidth is given by\cite{die83} $\gamma \! = \! \sqrt{ 8  E_{P}k_{B}T \ln 2}$ and therefore allows us to estimate $E_{P}$. For the dot in Fig.~1a, we get $E_{P} \! = \! 3.4$meV, which is comparable to the transition linewidth. We would therefore expect the transition to redshift by about the linewidth energy, if the polaron was formed in this dot. However, as seen in Fig.~1a (bottom), the recombination energy remains constant throughout the exciton lifetime. The small value of $E_P$ is not the only reason that the MP formation is not visible in this dot. As follows from the studies on bulk and 2D systems, the formation dynamics slows down substantially with the decrease of $x$ \cite{zay87,mac94} with $\tau_f$ exceeding $\tau_{X}$ for $x\! < \! 0.1$.\cite{mac94}

A different situation is encountered when $x$ is increased to $0.035$ -- Fig.~1b. Here, a redshift of the transition is clearly present throughout the entire exciton lifetime. The linewidth is increased to $\gamma \! = \!6$ meV allowing to estimate $E_{P} \! = \!9.4$ meV. Therefore, for this dot we are able to estimate $\tau_f$ by fitting the transition energy\cite{zay87,die95} with $E(t)\!=\!E_0 - E_{P}(1-\exp(-t/\tau_f))$. The fit is shown as a red line at the bottom of Fig.~1b and it yields $\tau_f\! = \! 3.6$ ns. Clearly, the MP formation is taking place, but it is interrupted by exciton recombination before equilibrium is reached. A similar behavior is found for QDs with $x\! = \! 0.01$ and a small volume, i.e. emitting in the high energy tail of the PL band. Squeezing the exciton wavefunction into a smaller volume yields larger $B_{\text{X}}$ in turn resulting in a larger $E_{P}$, which makes the initial stage of the MP formation visible during the exciton lifetime. 

As $x$ is increased, the MP formation further accelerates. As shown in Fig.~1c, for $x\! = \! 0.2$ we see a distinct redshift saturating about 300 ps after excitation. In this case, $E_P$ can be readily obtained from the experiment. For the QD in Fig.~1c, we get $E_{P}  \! = \!  13$ meV and $\tau_f \! = \! 110$ ps, while $\tau_{X} \! = \! 220$ ps.
Analogous transients are obtained for QDs with $x\! = \! 0.1$ with $\tau_f\approx 150$ ps. The decreasing $\tau_f$ values obtained for QDs with increasing $x$ agree very well with spin-spin relaxation times and MP formation times obtained for bulk and 2D systems.\cite{die95} We therefore conclude that the spin-spin interactions are responsible for the MP formation process also in QDs.

\begin{figure}[t]
  \includegraphics[angle=0,width=.45\textwidth]{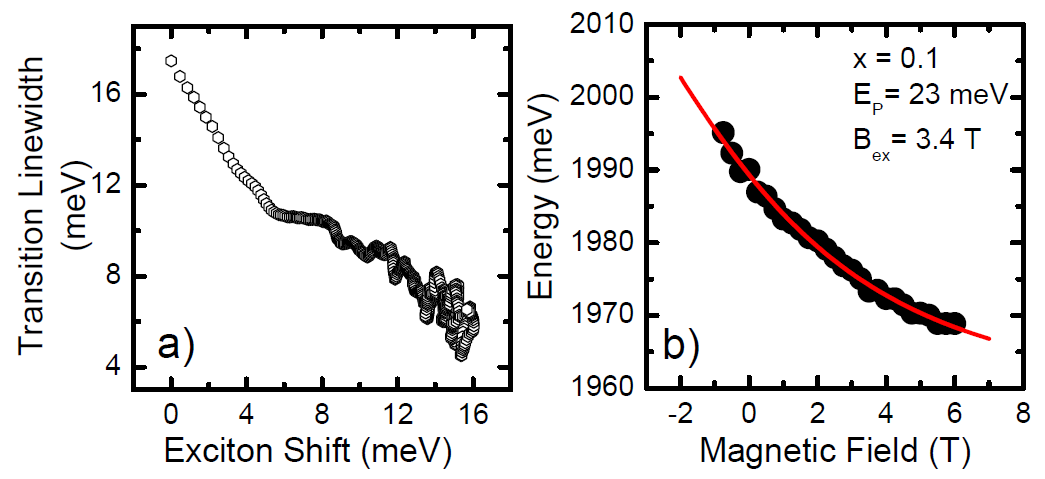}
  \caption{a) Transient PL linewidth plotted as a function of transient exciton energy for a QD with $x=0.1$. b) Points: PL energies of exciton recombination from a QD with $x=0.1$ as a function of external magnetic field. Line: Fitted Brillouin function (see text). }
  \label{ZL}
\end{figure}

The build-up of the spontaneous magnetization suppresses the thermal fluctuations of the Mn spins responsible for the broadening of the QD transitions. In Fig.~\ref{ZL}a, we plot $\gamma$ as a function of the exciton energy shift for a QD with $x$=0.1. Indeed, we find that the energy shift resulting from the MP build-up is accompanied by a substantial decrease of $\gamma$.

In the case of QDs in which the equilibrium is reached within $\tau_{X}$, we can obtain the MP parameters from cw spectroscopy. In order to estimate $B_{\text{X}}$, we measured cw PL as a function of external magnetic field $B_{0}$ applied in Faraday configuration. In Fig.~2b) we plot exciton transition energies from a QD with $x$=0.1 as a function of $B_{0}$. We fit these energies with  $E_{PL}(B) = \Delta E_{sat}(x)/2 B_S[g \mu_B B S/(k_B (T+T_0(x)))]$, where $B_S$ is a Brillouin function for a spin $S \! = \! 5/2$ and $\Delta E_{sat}$ and $T_0$ are phenomenological parameters taking into account the antiferromagnetic coupling of the Mn ions \cite{gaj94} and $k_B$ is the Boltzmann constant. Since the emission occurs after the equilibrium was reached, the Mn ion system experiences the sum of the exchange and external fields. Consequently, $E_{PL}(B) = E_{PL}(B_{0}+B_{\text{X}})$ allowing us to extract\cite{mak00} $B_{\text{X}}$ and $E_P=E_{PL}(0)-E_{PL}(B_{\text{X}})$. From the fit, we get $B_{\text{X}} \! = \! 3.4$ T and $E_P \! = \! 23$ meV, which corresponds to the localization volume spanning over roughly 1500 cation sites.

For $x\! < \! 0.02$ we can neglect antiferromagnetic couplings between the neighboring Mn spins and estimate the localization volume from the cw transition linewidth. For noninteracting Mn spins, regardless of the MP being formed $\gamma \! \approx \! xN_{0}(\beta-\alpha)\sqrt{35/48xN_{0}V}$. A survey of 20 dots gives us a mean linewidth of 3.3 meV, which corresponds to $V \! \approx \! 4500N_{0}^{-1}$. Assuming QD height of 2 nm, equal to the nominal thickness of the CdMnTe\ layer, we obtain the lateral size of the exciton wavefunction of $\sim$ 20 nm.

\begin{figure}[t]
  \includegraphics[angle=0,width=.45\textwidth]{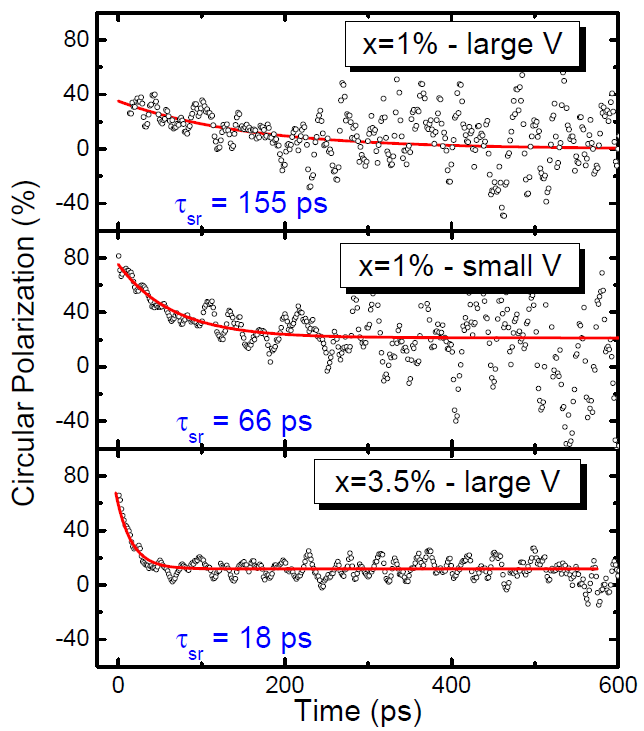}
  \caption{Temporal dependences of PL circular polarization degrees for QDs with 1\% Mn cations and a large volume (top), 1\% Mn cations and a small volume (middle), and 3.5\% Mn cations (bottom). $\tau_{\text{sr}}$ denotes the spin relaxation time evaluated from the monoexponential decays (lines).}
  \label{polarization}
\end{figure}

We distinguish between two scenarios of the MP formation: (i) the Mn spins align parallel to the exciton spin thus lowering the total energy of the system (this would lead to preferential orientation of MPs along the direction chosen by the circular polarization of light\cite{mac04}); (ii) the exciton adjusts its spin to a momentary magnetization fluctuation and subsequently amplifies it as the Mn spins align (this would lead to a loss of exciton spin polarization before the MP formation).
In order to identify the scenario occurring in the studied samples, we investigate the dynamics of exciton spin relaxation by analyzing the temporal dependence of the PL circular polarization $\varrho = (I^+-I^-)/(I^++I^-)$, where $I^{\pm}$ are intensities of PL signals co- and cross-polarized with respect to the excitation beam. For samples with $x \leq 0.035$, due to a small energy which needs to be relaxed to reach the emitting QD state or due to an interdot excitation transfer,\cite{kaz09}
the emission at short delays preserves a significant fraction of the laser polarization -- see Fig.~\ref{polarization}. This optical orientation is lost within the exciton spin relaxation time $\tau_{\text{sr}}$. We thus identify $\tau_{\text{sr}}$ with the decay time of $\varrho$. In Fig.~\ref{polarization}, we show $\varrho(t)$ for three QDs. We observe a distinct decrease of $\tau_{sr}$ with increasing Mn fluctuation induced exciton splitting, i.e.~going from top to bottom of Fig.~\ref{polarization} $\tau_{\text{sr}}$ decreases from 155 ps to 18 ps.

In the QDs studied here we have $\tau_{\text{sr}} \! \ll \! \tau_{f}$ -- compare Figs.~\ref{raw} and \ref{polarization}.
We therefore conclude that the exciton spin is relaxed before the formation of the polaron. The same conclusion was drawn for bulk and 2D systems.\cite{mer95} This formation scenario inhibits the optical orientation of MPs reported for an ensemble of smaller dots.\cite{mac04}

Remarkably, these spin relaxation times are much shorter than in the case of CdTe dots measured in finite magnetic fields. \cite{mac03,gor10bri} This acceleration is rather not due to an increase of phonon-induced spin relaxation rate with exciton Zeeman splitting\cite{Tsi03,Ros07,tsi10} generated by the fluctuating Mn spins, since our estimations show that the magnitude of this field is equivalent to only a few Tesla -- comparable with the fields in the cited experiments.
\cite{mac03,gor10bri} Alternatively, the exciton spin can be relaxed due to its interactions with the Mn spin bath. Quick fluctuations stemming from spin-spin interactions provide an efficient channel for both energy and spin relaxation. The latter originates from angular momentum conservation breaking by the anisotropic spin-spin interactions.\cite{die95} Thus the process that allows for creation of visible magnetic moment at larger $x$ might explain the nonconservation of the exciton spin at smaller $x$.


Our result do not address directly the dynamics of the decay of ferromagnetic order after the exciton recombination. We assume that this relaxation time is roughly equal to $\tau_f$.\cite{seu02,die95} On the other hand, studies on small \CdMnTe\ QDs with $x\! = \! 0.02$ revealed persistence of the magnetic order for as long as 100 $\mu$s.\cite{gur08} This long lasting spin memory manifested itself in the PL polarization independent of the excitation polarization. The transients shown in Fig. \ref{polarization} in fact display an offset of about 20\% in the long-delay value of $\varrho(t)$. We are, however, unable to conclude on its origin. Moreover, measurements performed on QDs with $x=0.035$, where the exchange field is substantial while  $\tau_f$ remains comparable to the laser repetition period, have not shown any signatures of long-living magnetization memory.


In summary, we have studied the dynamics of magnetic polaron formation in single \CdMnTe\ quantum dots. We found that the formation is accelerated with an increase of the Mn ion density, in accordance with the increasing strength of the spin-spin interactions responsible for the process. We found that the polaron is formed after the exciton relaxes its spin and the polaron orientation is related to random magnetization fluctuation which the exciton encounters upon creation. These conclusions are based on the analysis of exciton spin relaxation rates, which are several orders of magnitude larger than in nonmagnetic CdTe dots.


This research was supported by a Polish Ministry of Science and Education grant no. 0634/BH03/2007/33 and the Polonium Programme and by European Union within European
Regional Development Fund, through grant Innovative Economy (POIG.01.01.02-00-008/08 and POIG.01.03.01-00-159/08). {\L}C acknowledges support from the Homing programme of the Foundation for Polish Science and "FunDMS" Advanced Grant of the ERC within the "Ideas" 7th Framework Programme. We thank Tomasz Dietl for valuable discussions.

\end{document}